\documentclass[11pt]{article}
\usepackage{amssymb,amsthm,amsfonts,graphicx,multicol,mathrsfs}
\vspace{.2in}
\begin{document}

\pagestyle{empty}
\def\thefootnote{\fnsymbol{footnote}}

\vspace*{3cm}

\centerline{\Large \bf  Up Quark Masses from Down Quark Masses}  
\vspace*{2cm} 

\centerline{ \large  I. Masina$^a$ and C.A. Savoy$^b$ } 
\vskip 1cm

\begin{center}
\textit{a) Centro Studi e Ricerche "E. Fermi", Via Panisperna 89/A, Roma,\\
and INFN, Sezione di Roma, Italy}
\end{center} 
\begin{center}
\textit{b) Service de Physique Th\'eorique, CEA-Saclay, Gif-sur-Yvette, France}
\end{center}

\vskip 2cm
\begin{center}{\bf Abstract} \end{center}

The quark and charged lepton masses and the angles and phase of the CKM   
mixing matrix are nicely reproduced in a model 
which assumes $SU(3) \otimes SU(3)$ flavour symmetry 
broken by the v.e.v.'s of fields in its bi-fundamental representation. 
The relations among the quark mass eigenvalues, 
$m_{u} / m_{c} \approx m_{c} / m_{t} \approx 
m^{2}_{d} / m^{2}_{s} \approx m^{2}_{s} / m^{2}_{b} \approx 
\Lambda^{2}_{\mathrm{GUT}} / M^{2}_{\mathrm{Pl}}$, follow from  
the broken flavour symmetry. 
Large $\tan \beta$ is required which 
also provides the best fits to data for the obtained textures. Lepton-quark  
grandunification with a field that breaks both $SU(5)$ and the flavour group correctly 
extends the predictions to the charged lepton masses. The seesaw extension of the 
model to the neutrino sector predicts a Majorana mass matrix quadratically hierarchical 
as compared to the neutrino Dirac mass matrix, naturally yielding large mixings and 
low mass hierarchy for neutrinos.

\newpage
\setcounter{page}{1}
\pagestyle{plain}
\def\thefootnote{\arabic{footnote}}
\setcounter{footnote}{0}

In a bottom-up approach to the flavour problem, the first task is to decode the variety
of experimental data on fermion mass ratios and mixings. 
By translating that information into simple mass matrix textures 
that reproduce the empirical relations one makes progress towards the interpretation 
of these textures in terms of flavour symmetries and their breakings. 
In spite of the many interesting textures and models present in the literature, 
the flavour issue still remains a totally open problem. 
We still ignore why $m_{t} \gg m_{b}$ while $m_{u} < m_{d}$.

Since the first suggestion of a relation between the Cabibbo angle and quark 
mass ratios \cite{gatto}, many models were built where the angles of the 
CKM mixing matrix are expressed in terms of the quark mass eigenvalues 
(in practice angles are better known than light quark masses). 
Many textures have been designed along these lines, in particular,
those with a maximal number of vanishing matrix elements 
which are now excluded by the experiments \cite{georgi, excluded}. 
More recent ones \cite{rrrv} improve the fit to the data by 
enriching the textures, however these are not quite consistent 
with the latest data \cite{mas-sa}.  
Neutrino oscillation experiments have prompted a lot of work on the 
analogous approach to the lepton mass matrices including solutions compatible
with lepton-quark (grand-)unification \cite{rom}. The considerable progress 
obtained in the last years in measurements of the quark  mixings - all quite 
consistent with the Standard Model (SM) description of flavour changing and 
CP violation effects \cite{fit} - as well as in studies of neutrino oscillations, 
implies a much stronger selection of allowed textures. 

One obvious difficulty in the definition of the mass textures is their dependence
on the assumed basis for the fermions since only the family mixings inside
the weak doublets of quarks or leptons is observable without physics beyond 
the SM. This allows for several different patterns of mass textures even after 
the number of free parameters are reduced by theoretical assumptions or educated 
guess. This fact notwithstanding, these efforts are an important step in the quest
for the symmetries underlying the flavour theory which should naturally ensure
these relations. Because of the hierarchical nature of the charged fermion masses
and observed mixings, the flavour theory must also contain one or more small
parameters whose existence would be natural only if protected by spontaneously 
broken flavour symmetry.

In models based on abelian symmetries, the different scales present in the
mass matrices are associated to powers of the small parameters defined by 
the choice of the fermion abelian charges, all their $O(1)$ complex pre-factors 
remaining arbitrary \cite{ab}. This intrinsically limits the predictivity 
of these models and their selection by the experimental progress. 
Instead, non-abelian flavour symmetries potentially establish exact relations 
and are more constrained but the overall hierarchies require more than one small 
parameter and more involved symmetry breaking schemes \cite{non-ab}. 
In this paper we develop a novel approach to these issues based on the following 
argumentation.

In a recent publication \cite{mas-sa}, we have built fermion mass matrices 
by the identification of a few characteristic features of the mixing angles and phases
and their implementation via simple mechanisms and associated textures. Some 
ingredients were already present in the literature (as referred to in ref. \cite{mas-sa}), 
but were combined to bring forth new textures for $\mathbf{m}_{\mathrm{up}}$ and 
$\mathbf{m}_{\mathrm{down}}$ following some observations outlined below. 
This resulted in  textures with five free parameters: two that implement the 
double seesaw-like texture of $\mathbf{m}_{\mathrm{up}}$, often advocated in the 
literature, plus a necessary, smaller parameter to improve the fit of $m_{u}$ to data; 
and two that define a new texture for $\mathbf{m}_{\mathrm{down}}$. 
CP violation is introduced by requiring the so-called maximal CP violation between 
two families (namely a phase $i$) \cite{maxCP}.  
The fit nicely reproduces the masses and the unitarity triangle 
within the relatively small experimental uncertainties. 

\begin{figure}[!t]
\vskip 1. cm
\centerline{\includegraphics{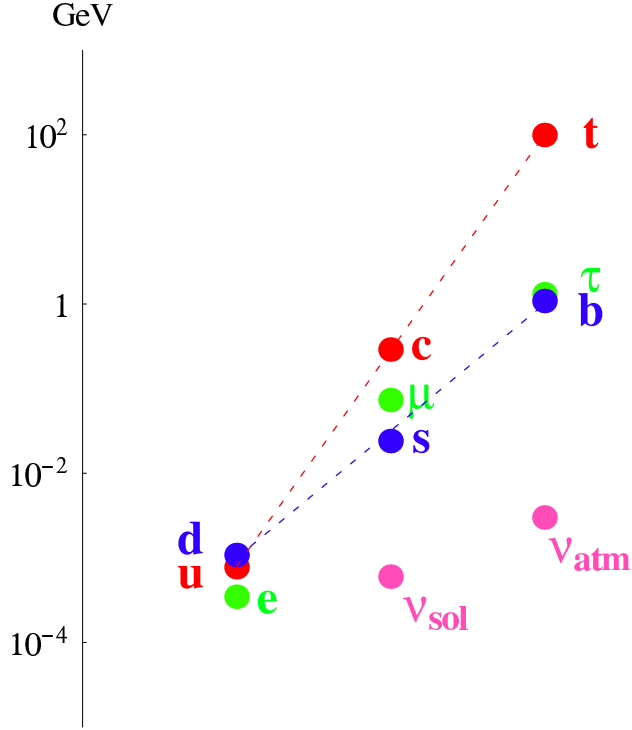}} 
\caption{Fermion masses at the scale $M_{\mathrm{GUT}}$ for $\tan \beta=30$.
The neutrino mass scales $\sqrt{\Delta m^2_{sol}}$ and $\sqrt{\Delta m^2_{atm}}$ 
have been rescaled by a factor $10^7$.}
\label{fig1}
\vskip 1. cm
\end{figure}

In order to further reduce the number of parameters, so to get more insight into the
underlying flavour symmetry, further relations about mass ratios in and between
$\mathbf{m}_{\mathrm{up}}$ and $\mathbf{m}_{\mathrm{down}}$ are needed
besides those among  mixing angles and mass ratios implicit in these textures. There
are indeed other intriguing relations in the hierarchies of  the $\mathbf{m}_{\mathrm{up}}$
and $\mathbf{m}_{\mathrm{down}}$  eigenvalues that are latent guidelines in flavour 
model  building, namely, the following approximate relations:
\begin{equation}
\frac{m_{u}}{m_{c}} \approx \frac{m_{c}}{m_{t}} \approx 
\frac{m^{2}_{d}}{m^{2}_{s}} \approx \frac{m^{2}_{s}}
{m^{2}_{b}} = O\left(  \frac{M^{2}_{\mathrm{GUT}}}
{M^{2}_{\mathrm{Pl}}}\right) ~~. \label{magic}
\end{equation}
They are better realized for the values of the quark masses run up to the GUT scale
$M_{\mathrm{GUT}}$ in a supersymmetric framework, shown in fig. 1, which 
also makes the last relation more suggestive. 

The general strategy in the quest for hidden symmetries in the fermion masses and
mixings assumes that they are defined by appropriate effective operators after the
flavour symmetry breaking fields are replaced by their v.e.v's. Therefore one can 
write:
\begin{eqnarray}
\mathbf{m}_{\mathrm{down}} = \sum _{i} \mathbf{X}_{i} v \cos\beta   
\label{linear}
\end{eqnarray}
where the $\mathbf{X}_{i}$ matrices are functions of v.e.v.'s associated to the various
scales needed to define the fermion masses in units of the cutoff\footnote{In this paper 
the cutoff is the Planck scale and the $\mathbf{X}_{i}$ are the most general ones 
allowed by the hidden symmetries. In supersymmetric models 
with a cutoff scale much below $M_{\mathrm{Pl}}$ it makes sense to select some
$\mathbf{X}_{i}$ by a choice of states to be integrated out.}, 
and $v = 174~\mathrm{GeV}$. The relations (\ref{magic}) 
suggest the following expression: $\mathbf{m}_{\mathrm{up}} = 
\sum _{i,j}b_{ij} \mathbf{X}_{i}^\dagger \mathbf{X}_{j} v\sin\beta $, which, for 
$b_{ij} = 1$, trivially satisfies (\ref{magic}) with vanishing mixings. 
Then by a choice of $O(1)$ numbers $b_{ij}$ the CKM angles and phase could be introduced.
Of course, the program makes sense only if the number of free parameters is low enough. 
However, in a supersymmetric theory $\mathbf{m}_{\mathrm{up}}$ have to be 
holomorphic in the fields $\mathbf{X}_{i}$, suggesting to replace that expression by
\begin{equation}
\mathbf{m}_{\mathrm{up}} = \sum _{i,j} b_{ij} \mathbf{X}_{i}^T  \mathbf{X}_{j}
 v\sin\beta  ~.\label{quad}
\end{equation}
Let us define the relation corresponding to (\ref{linear}) and (\ref{quad}) as 
$\mathbf{m}_{\mathrm{up}} = 
\mathbf{m}^{T}_{\mathrm{down}}\circ \mathbf{m}_{\mathrm{down}}$. In the presence 
of a flavour symmetry, $\mathbf{m}_{\mathrm{up}}$ correctly transforms only under 
$O(3)$ transformations of the right-handed down quarks. In spite of the fact that 
$\mathbf{m}_{\mathrm{up}}=\mathbf{m}^{T}_{\mathrm{down}}\circ \mathbf{m}_{\mathrm{down}}$ 
alone does not ensure the relations (\ref{magic}), we present in this letter a realistic model 
that does\footnote{For recent work addressing the relations (\ref{magic}) in a different 
approach see, \textit{e.g.}, Ref.\cite{nesti}.}.

We seek out a flavour model realizing both the textures of \cite{mas-sa} 
and the hierarchies in (\ref{magic}) in the framework of effective supergravity and
grandunification. We propose a model with $SU(5)\otimes SO(3)\otimes SU(3)$
symmetry under which fermions transform in  one $(\underline{\bar{5}},\,
\underline{3},\, \underline{1})$ plus  one $(\underline{10},\,\underline{1},\, 
\underline{3})$  (right handed neutrinos would require another $SU(3)$ factor), and
the electroweak symmetry breaking Higgs doublets come from a $(\underline{\bar{5}}
\oplus \underline{5},\,\underline{1},\, \underline{1})$. This flavour symmetry is broken by 
the v.e.v.'s of three fields, all transforming as the bifundamental representation of the 
flavour symmetry. Later on the flavour group is upgraded to $SU(3)\otimes SU(3)$,
where one factor is broken to $SO(3)$ by a field in its symmetric represention.

The Yukawa couplings in $\mathbf{m}_{\mathrm{up}}$ and 
$\mathbf{m}_{\mathrm{down}}$  are obtained from $SU(5)\otimes SU(3)
\otimes SU(3)$ invariant higher-dimension operators  after these fields are 
substituted with their v.e.v.'s. The latter follow from  field equations designed 
to reproduce the required textures.  There are only two
small parameters, the two ratios between v.e.v.'s, one actually corresponding
to ${M_{\mathrm{GUT}}}/{M_{\mathrm{Pl}}}$. Consequently, the 
three small parameters in $\mathbf{m}_{\mathrm{up}}$ are related to the 
two fitted ones  in $\mathbf{m}_{\mathrm{down}}$, up to $O(1)$ factors
associated to couplings in the effective supergravity. The relation
$\mathbf{m}_{\mathrm{up}} = \mathbf{m}^{T}_{\mathrm{down}}
\circ \mathbf{m}_{\mathrm{down}}$ is realized.

Therefore, this model  explains why the hierarchy in $\mathbf{m}_{\mathrm{up}}$ 
is approximately quadratic with respect to that in $\mathbf{m}_{\mathrm{down}}$: 
it results from both the chiral-like flavour symmetry and the direction of the v.e.v's 
also required by the successful textures. The charged lepton mass eigenvalues are
simply obtained \`a la Georgi-Jarlskog by promoting one of the flavour breaking 
fields to be the  $\underline{75}$ or the $\underline{24}$ that breaks the 
$SU(5)$ GUT symmetry.  Interestingly enough, this is also required to explain one
small parameter, namely the ratio $m_{s}/m_{b}$, as well as to optimize 
$\mathbf{m}_{\mathrm{up}}$. 

Paradoxically, the setup also provides a solution for the large mixing and 
low hierarchy that characterizes the neutrino effective mass matrix even if the 
seesaw Yukawa couplings are  hierarchical like the quark and charged lepton ones. 
Indeed, with a further flavour $SU(3)$ factor associated to right-handed neutrinos, 
their Majorana mass matrix will result quadratically hierarchical as compared 
to the Dirac mass matrix, $\mathbf{m}_{\mathrm{M}} = 
\mathbf{m}_{\mathrm{D}}\circ \mathbf{m}_{\mathrm{D}}^T$, analogously
to the quark sector mass matrices.  By a strong compensation of hierarchies, the 
seesaw mechanism then yields an effective neutrino mass matrix consistent with 
the experimental one.

\vskip 12pt
\underline{\textit{Mass Matrix Textures}}

Let us first summarize the steps that lead to the mass textures in \cite{mas-sa}. 
We first concentrate on the quark sector and write the mass matrices as follows:
\begin{equation}
\mathbf{m}_{\mathrm{up}} = U_{u_{R}} \mathrm{diag}(m_{u},m_{c},m_{t})
U_{u_{L}}^{\dag }, \qquad 
\mathbf{m}_{\mathrm{down}} = 
U_{d_{R}} \mathrm{diag}(m_{d},m_{s},m_{b})U_{d_{L}}^{\dag },
\end{equation}
so that the CKM matrix is $U^{CKM}={U_{u_L}}^\dagger ~{U_{d_L}}$. The 
unitary  matrices are written as: 
\begin{equation}
U = e^{i\Phi}R(\theta_{23})\Gamma_{\delta} R(\theta_{13})
\Gamma^{\dagger}_{\delta} R(\theta_{12}) e^{i\Phi '}
\end{equation}
where $\Phi = \mathrm{diag}(\phi_{i})$, $\Phi' = \mathrm{diag}(\phi_{i}')$,
$\Gamma_{\delta}= \mathrm{diag}(1,1,e^{i\delta})$ and $R(\theta_{ij})$ is
a rotation in the $(i,j)$ plane.  Only the phases $\phi_{ij} = \phi^{d_{L}}_{i}-
\phi^{u_{L}}_{i}-\phi^{d_{L}}_{j}+\phi^{u_{L}}_{j}$ are relevant for
the CKM matrix.

Our basic assumptions are based on two simple facts. Firstly, the relation 
$|U^{CKM}_{ub}| = O  ( | U^{CKM}_{us} U^{CKM}_{cb} | )$ suggests  that  
$\theta_{13}$ in $U^{CKM}$ results from the commutation between  the other 
rotations. Accordingly, we have assumed in \cite{mas-sa} that  $\theta^{u_{L}}_{13} 
=  \theta^{d_{L}}_{13} = 0$. Secondly, it then follows that the unitarity triangle 
angle $\alpha =\phi_{12}$ and experimentally it is consistent with $\pi / 2$. 
Instead, $\phi_{23}$ has to be small. Therefore we assume one and only 
one phase $\phi_{12} = \pi / 2$   in  $\mathbf{m}_{\mathrm{down}}$ and 
a real $\mathbf{m}_{\mathrm{up}}$. The complete analysis in 
ref. \cite{mas-sa} leads to the following textures:
\begin{eqnarray}
\mathbf{m}_{\mathrm{up}} = m_{t}\left(\matrix{c \lambda^8 
& a \lambda^6 & 0 \cr  -a\lambda^6 & c \lambda^8 & -b \lambda^2 \cr 
 0 & b \lambda^2 & 1   } \right)   \qquad   
\mathbf{m}_{\mathrm{down}} = m_{b} \left(\matrix{0 & g \lambda^{3} 
& 0 \cr g \lambda^{3}&i f \lambda^2  & 0 \cr  0 & f \lambda^2  & 1 } 
\right)  \label{textures}
\end{eqnarray}
up to  unobservable unitary transformations\footnote{Actually some matrix 
elements in the textures displayed in \cite{mas-sa} have opposite signs with 
respect to (\ref{textures}). Both choices give the same results for the fitted 
observables but only the textures (\ref{textures}) are consistent with the model 
discussed in this paper.}. The Wolfenstein approximation for $U^{CKM}$ 
suggests to introduce  in (\ref{textures}) powers of $\lambda = \sin \theta_{C}$ 
to roughly characterize the magnitude of the mass matrix elements as well.

In (\ref{textures}), $\mathbf{m}_{\mathrm{up}}$ has the double seesaw texture 
with the additional parameter $c \lambda^{8}$ to shift $m_{u}$ down 
and obtain a good fit to all data\footnote{While in \cite{mas-sa} $c\lambda^{8}$ 
appears only in $(\mathbf{m}_{\mathrm{up}})_{11}$, in the present 
model it also appears in $(\mathbf{m}_{\mathrm{up}})_{22}$  and negligibly 
affects the results.}, while $\mathbf{m}_{\mathrm{down}}$ has only two 
parameters and one single phase, $i$, which yields $\alpha \approx  \pi/2$. 
The fit to the ten experimental observables is quite satisfactory. The mass 
matrices are defined at the GUT scale and the couplings are run down to the 
electroweak scale for comparison with data, which introduces a dependence
on $\tan \beta$. The fit is better for larger values of $\tan \beta$. The results for
$\tan\beta = 45$ are as follows: 
\begin{equation}
f= 0.33 ,\quad g= 0.32 ,  \quad a= 1.77 ,\quad b= 1.01 ,\quad c= -3.6 ~.\label{fit}
\end{equation}
For more details we refer to \cite{mas-sa}.

The matrices in (\ref{textures}) are the simplest realization of our assumptions
up to unitary transformations that  do not modify the measured observables, 
provided they do not introduce new parameters. But we stick to the textures in 
(\ref{textures}) since it has a nice explanation in terms of a flavour model
that we now turn to discuss. In \cite{mas-sa} the charged  lepton mass matrix,
$\mathbf{m}_{\mathrm{\ell}}$, was obtained by using the $SU(5)$ relation
$\mathbf{m}_{\mathrm{\ell}}$ = $\mathbf{m}^{T}_{\mathrm{down}}$,
but assuming that the coupling $f$ transforms as an element of a 
$\underline{45}$ of $SU(5)$, so that
\begin{equation}
\mathbf{m}_{\mathrm{\ell}} = m_{b} \left(\matrix{0 & g \lambda^{3} 
& 0 \cr g \lambda^{3}& -3i f \lambda^2  & -3f \lambda^2 \cr  0 & 0   & 1 } 
\right) . \label{mlept}
\end{equation}
This reasonably fits  the charged lepton masses, realizing the relations:
$m_{e}\approx m_{d}/3,\,m_{\mu}\approx 3m_{s},\, m_{\tau}\approx 
m_{b},\, $  at  $M_{\mathrm{GUT}}$.

\vskip 12pt
\underline{\textit{A Flavour Model: Quarks}}

Let us begin with  the quark sector and  subsequently extend the results to the 
lepton one. The maximum flavour symmetry of the gauge interactions in a $SU(5)$
GUT is a chiral $SU(3)\otimes SU(3)$ where the two factors act on the three 
$\underline{\bar{5}}$'s and three $\underline{10}$'s,
respectively. As already stated, we first propose a model with the $SO(3) 
\otimes SU(3)$ subgroup as the overall (possibly gauged) flavour symmetry. 
Therefore, $\mathbf{m}_{\mathrm{down}}$ belongs to the 
$(\underline{\bar{5}}\otimes \underline{10})$  representation of $SU(5)$ 
and the $(\underline{3}, \underline{\bar{3}})$ of the flavour group, 
while $\mathbf{m}_{\mathrm{up}}$ 
transforms in the $(\underline{10}\otimes \underline{10})$ and  in 
the $(\underline{1}, \underline{\bar{3}}\otimes \underline{\bar{3}})$,  respectively.
Fermion masses are assumed to derive from the effective Yukawa couplings to the 
electroweak Higgs doublets in a $\underline{\bar{5}}\oplus \underline{5}$, 
invariant under the flavour symmetry, with the usual doublet-triplet splitting. 

These Yukawa couplings are functions of the flavour symmetry breaking v.e.v.'s  
defined by the allowed $SU(5)\otimes SO(3)\otimes SU(3)$ operators.  To realize 
the hierarchy and texture of $\mathbf{m}_{\mathrm{down}}$ in (\ref{textures}) 
the corresponding fields must be in the $(\underline{3} ,\,  \underline{\bar{3}})$ and 
we need three of them with v.e.v.'s:
\begin{eqnarray}
\mathbf{P} = \left(\matrix{0 & 0 & 0 \cr 0 & 0 & 0 \cr  0 & 0 & 1}
\right) \Lambda , \quad 
\mathbf{F} = \left(\matrix{0 & 0 & 0 \cr   0& i  & 0 \cr  0 & 1  & 0 } 
\right)f \lambda^2\Lambda , \quad 
\mathbf{G} = \left(\matrix{0 & 1 & 0 \cr  1 & 0 & 0 \cr 0 & 0 & 0} \right)  
 g \lambda^{3}\Lambda ,  \label{vevs}
\end{eqnarray}
where the $O(1)$ parameters $g$ and $f$ and the flavour symmetry breaking scale 
$\Lambda$ are to be fixed by fitting $\mathbf{m}_{\mathrm{down}}$.
These v.e.v.'s  are the solution (up to complexified $SO(3)\otimes SU(3)$ 
transformations) of the following analytic field equations,
\begin{eqnarray}
\mathbf{P}^{T}\mathbf{P}&=&\Lambda \mathbf{P} 
\qquad  \quad  \mathrm{Tr}\mathbf{P} = \Lambda \qquad  \qquad 
\mathbf{P}\mathbf{F}^{T} = 0  \nonumber\\ 
\mathbf{F}^{T}\mathbf{F}&=&0 \qquad   \qquad 
\mathbf{P}^{T}\mathbf{G} = 0 \qquad  \qquad \mathbf{F}^{T}\mathbf{G} 
+ \mathbf{G}^{T}\mathbf{F} =  f \lambda^2\Lambda \mathbf{G} ~~.
\label{fieldeqs}
\end{eqnarray}
At the scale $\Lambda$,  the flavour symmetry is broken
to $SO(2)\otimes U(2)$ by $\mathbf{P}$, which acts as a projector 
of the heavy family. The field equations can be implemented in the superpotential 
(invariant under the complexified flavour group) by introducing Lagrange 
multiplier fields, although this method could look awkward. The construction 
of a more satisfactory superpotential will be presented elsewhere (also because 
it would depend on some options defined below) and we concentrate in the 
following on the consequences of (\ref{vevs}) for the fermion masses.

Since the natural cutoff of the supersymmetric GUT  is the Planck mass 
$M_{\mathrm{Pl}}$, $\mathbf{m}_{\mathrm{down}}$ can be written as
\begin{eqnarray}
\mathbf{m}_{\mathrm{down}} = \frac{\mathbf{P} + \mathbf{F} 
+ \mathbf{G}}{M_{\mathrm{Pl}}}~ v\cos\beta  ~~, 
\label{mdown}
\end{eqnarray}
where $O(1)$ real coefficients have been absorbed by a redefinition of  $f$, $g$ 
and $\Lambda$. In particular, $m_{b} = \Lambda v \cos\beta /M_{\mathrm{Pl}} $. 

Let us derive its consequences for $\mathbf{m}_{\mathrm{up}}$ with the same 
set of v.e.v.'s \footnote{In the previous version of this paper, with flavour group 
$O(3)\otimes O(3)$, $\mathbf{m}_{\mathrm{up}}$ would have a flavour singlet 
component implying an equal mass contribution for all the up-quarks. This 
cannot be forbidden by the assumption of discrete symmetries as suggested 
there. We thank Z. Berezhiani for calling our attention to this problem. }. 
The lowest order $SO(3)\otimes SU(3)$ invariant  operators are quadratic 
in $\mathbf{P}$, $\mathbf{F}$ and $\mathbf{G}$. Combining them into 
products that transform as  $(\underline{1},\, \underline{\bar{3}} \otimes 
\underline{\bar{3}})$ and replacing the solutions (\ref{fieldeqs}) yields the 
general expression:
\begin{eqnarray}
\mathbf{m}_{\mathrm{up}} = \frac{v \sin \beta}{ M^2_{\mathrm{Pl}} } 
\left( p (\mathbf{P}^{T}\mathbf{P}) + 
q(\mathbf{P}^{T}\mathbf{F} - \mathbf{F}^{T}\mathbf{P}) +
q'(\mathbf{P}^{T}\mathbf{F} + \mathbf{F}^{T}\mathbf{P}) + \right.\nonumber \\
\left. r(\mathbf{F}^{T}\mathbf{G} - \mathbf{G}^{T}\mathbf{F}) + 
r'(\mathbf{F}^{T}\mathbf{G} + \mathbf{G}^{T}\mathbf{F}) + 
s(\mathbf{G}^{T}\mathbf{G}) \right) 
\label{mup}
\end{eqnarray}
where the couplings $p$, $q$, $q'$, $r$, $r'$ and $s$, of the corresponding invariant 
operators in the superpotential are expected to be $O(1)$. With $q'=r'=0$ (to be 
discussed below), the expression in  (\ref{mup}) nicely matches the texture for  
$\mathbf{m}_{\mathrm{up}}$ in (\ref{textures}). By comparison we get 
$q/p=b/f=3.1$, $r/p= ia\lambda /fg=3.8i$, and $s/p=c\lambda^{2}/g^{2} =-1.9$
with the fitted values in (\ref{fit}) for $\tan\beta = 45$. Of course, the fact 
that these numbers come out  $O(1)$ is a crucial check of the model. Also, notice 
the relevance of the condition $\mathbf{F}^{T}\mathbf{F}=0$ to obtain the 
right texture.

From (\ref{mdown}) and (\ref{mup}), with the masses at the scale $\Lambda$: 
\begin{equation}
\frac{m_{t}}{v\sin \beta} = \frac{p\Lambda^{2}}{M_{\mathrm{Pl}}^{2}} 
\, , \quad \frac{m_{b}}{v\cos \beta} = 
\frac{\Lambda}{M_{\mathrm{Pl}}} \quad \Longrightarrow \quad
\Lambda \approx  \frac{0.8~M_{\mathrm{Pl}}}{\sqrt{p}} \,, \quad
\sqrt{p} \approx  \frac{1.2 ~m_{t}}{\tan \beta ~ m_{b}}~. \label{Lambda}
\end{equation}
Hence $\Lambda = O (M_{\mathrm{Pl}})$, meaning a first breaking of the
flavour symmetry close to the cutoff $M_{\mathrm{Pl}}$ and involving 
supergravity effects.  The magnitude of the three sequential flavour symmetry
breakings are $O(M_{\mathrm{Pl}})$ in $\mathbf{P}$,  
$O(\lambda^2 M_{\mathrm{Pl}})$ in $\mathbf{F}$,  
$O(\lambda^3M_{\mathrm{Pl}})$ in $\mathbf{G}$. 
With only two small parameters,
the model defines a relationship between $\mathbf{m}_{\mathrm{up}}$ 
and $\mathbf{m}_{\mathrm{down}}$ that  nicely reproduces the much 
stronger hierarchy of the eigenvalues of the former 
as well as the relative mixings and phase in the CKM matrix. 
Because $\Lambda = O (M_{\mathrm{Pl}})$, operators with higher powers of 
$\mathbf{P}/M_{\mathrm{Pl}}$ must be included in $\mathbf{m}_{\mathrm{down}}$ 
and $\mathbf{m}_{\mathrm{up}}$. With the fields above it is not possible to
write any new relevant operator that does no vanish with the assumed 
v.e.v.'s\footnote{Actually, the realization of the field equations would 
presumably require more fields, e.g., a field transforming as the conjugate
of $\mathbf{P} $. Since $\mathbf{P}/\Lambda$ is a projector, any polynomial
$\phi (\mathbf{P}/M_{\mathrm{Pl}}) = \phi (\Lambda / M_{\mathrm{Pl}})
\mathbf{P}/\Lambda$ so that the mass matrix textures would be mildly affected. }. 

We still have to naturally enforce $q'= r' =0$ or, equivalently, the vanishing 
of the corresponding symmetric operators in (\ref{mup}). As a matter of 
fact, this can be obtained from a simple assumption which turns out to be also 
required to fit the charged lepton spectrum (altogether this means a prediction). 
Indeed, let us take the fields in $\mathbf{F}$  to transform  under $SU(5)$ in a 
$\underline{75}$. Taking into account its product with the Higgs $\underline{5}$,
the two symmetric operators involving $\mathbf{F}$ in (\ref{mup}) transform
as  a $\underline{50}$ which has no colour singlet, e.w. doublet, so that they
do not contribute to $\mathbf{m}_{\mathrm{up}}$ as required. Correspondingly, 
in (\ref{mdown}) the term  $\mathbf{F}$ transforms as  the $\underline{\bar{45}}$. 
As an alternative, if $\mathbf{F}$ transforms as a $\underline{24}$, requiring 
that the its product with the Higgs $\underline{5}$ transforms as a $\underline{45}$, 
leads to the same consequences. In a sense, this assumption for the effective 
coupling is natural in the effective theory because it can be realized by choosing 
the states that are integrated out.

Consistently, we take the $\mathbf{F}$ v.e.v. for the $SU(5)$ breaking,
keeping in mind that an $O(1)$ coefficient has been absorbed in its 
definition in (\ref{mdown}). This defines the GUT scale as:
\begin{equation}
\Lambda_{\mathrm{GUT}}= O\left(  f\lambda^{2} \right) \Lambda 
= O\left(  10^{-2}\right)\, M_{\mathrm{Pl}} \label{GUT}
\end{equation}
with $f$ given in (\ref{fit}), which is quite consistent with the gauge 
coupling unification scale. Hence one  small parameter, $f\lambda^{2}$, 
is naturally related to  the  $\Lambda_{\mathrm{GUT}}$ scale. 
It remains one  parameter $g\lambda /f \approx \theta_{C}$ to be explained. 
The other four parameters are $O(1)$ unknown coefficients of the higher 
dimension operators in the  effective supergravity theory.

By choosing the field $\mathbf{F}$ to transform as a $(\underline{3},\, 
\underline{\bar{3}})$ of the flavour groups amounts to have nine $\underline{24}$'s
or nine $\underline{75}$'s of $SU(5)$. The model framework requires the
GUT gauge couplings to remain perturbative  at least up to the Planck scale,
which allows for five or six $\underline{24}$'s or one $\underline{75}$,
at most. Since the $\mathbf{P}$ projections suggest that three $\mathbf{F}$ 
components get masses of $O(M_{\mathrm{Pl}})$,  it seems consistent with 
perturbativity to assume $\mathbf{F}$ to transform as a $\underline{24}$.  
In the case of the $\underline{75}$, one must modify the model by writing 
$\mathbf{F}$ as the direct product of two matrices of fields: an $SU(5)$ 
singlet  $\mathbf{Q}$ in a bi-fundamental of the flavour group whose v.e.v. is 
$O(\Lambda)$ and analogous to $\mathbf{F}$ in eq. (\ref{vevs}), and 
a flavour singlet $\mathbf{V}$ in a $\underline{75}$ that breaks $SU(5)$ 
at the scale $f \lambda^{2}M_{\mathrm{Pl}}$. A discrete symmetry would 
constrain them to couple up in the mass matrices just as $\mathbf{F}$.  
  
As already noticed, the $SO(3)\otimes SU(3)$ flavour symmetry considered 
up to now can be explained by starting from  the more natural one, 
$SU(3)\otimes SU(3)$ with the fields  $\mathbf{P}\, ,  \mathbf{F} $ and 
$\mathbf{G}$ in the $(\underline{\bar{3}}\, , \underline{\bar{3}})$ and adding
a field $\mathbf{S}$ in the  $( \underline{6}\, ,1)$ representation;  
the operators in (\ref{mup}) now contain  $\mathbf{S}$ insertions:
$\mathbf{P}^T \mathbf{S}\mathbf{P}$, $\mathbf{F}^T \mathbf{S} \mathbf{P}$, \ldots  
When $\mathbf{S}$ gets an $SO(3)$ invariant v.e.v., 
diag$(1,1,1)\, \mathrm{O}(M_{\mathrm{Pl}})$,  the model discussed before 
is obtained.

Notice that all coefficients in (\ref{mup}) are real with the exception of $r$, 
with the phase $i$ needed  for $\mathbf{m}_{\mathrm{up}}$  to be real 
as in (\ref{textures}). In this way, the CP violation has been introduced by hand
- although it is nicely consistent with a maximal CP violation. Introducing the 
CKM phase through spontaneous CP symmetry breaking is a difficult problem 
by itself, specially in the context of a flavour theory with a reduced number of
fields. Interestingly, there is a simple mechanism to get spontaneous breaking 
of the CP symmetry in the context of the $SU(3)\otimes SU(3)$ flavour symmetry, 
but in pratice it does not work. Indeed, introducing  CP phases at the level of the
$SU(3 ) \rightarrow SO(3)$ breaking, 
$\mathbf{S}= \exp (i\mathrm{diag}(\alpha , \beta , 0))$, 
with $\beta \neq 0$ the important relation $\mathbf{F}^T\mathbf{SF}=0$ is lost,
while for $\alpha \neq 0$ the CP violation comes out wrong.

\vskip 12pt
\underline{\textit{Lepton-Quark Unification}}

The $SU(5)$ symmetry  relates $\mathbf{m}_{\mathrm{\ell }}$ to 
$\mathbf{m}^{T}_{\mathrm{down}}$  but the precise relation 
depends on insertions of $SU(5)$ breaking fields in the effective 
mass matrices. In \cite{mas-sa} a simple generalization of the 
Georgi-Jarlskog mechanism has  been proposed in order to  ensure the 
relations $m_{\mu}\approx 3m_{s}$ and $m_{e} m_{\mu}\approx 
m_{d}m_{s}$. It amounts to make the product of $\mathbf{F}$ and the 
electroweak Higgs  $\underline{\bar{5}}$ to transform as a
$\underline{\bar{45}}$. This is just what has been imposed above from 
the study of the quark sector. Thus the charged lepton mass matrix becomes 
a prediction of the model that reads:
\begin{eqnarray}
\mathbf{m}_{\mathrm{\ell}} =  \frac{\mathbf{P}^T -3 \mathbf{F}^T 
+ \mathbf{G}^T}{M_{\mathrm{Pl}}}  \, v\cos\beta \, .  
\label{mellnew}
\end{eqnarray}
It correctly accounts for the charged lepton mass eigenvalues 
within the precision appropriate to the aim of this paper \cite{mas-sa}. 

The mixing angles come out relatively small but, only by coupling this 
model to a neutrino mass generation mechanism, the prediction 
for $U_{\mathrm{MNS}}$ could be tested. Indeed, the observable mixing 
angles in neutrino oscillations are defined by the transformation 
$U_{\mathrm{MNS}}$ between the bases where the charged and neutral 
leptons in the $\bar{5}$'s are mass eigenstates respectively. Since in the 
basis chosen here the charged lepton angles are small\footnote{The 
alternative texture presented in Ref.\cite{mas-sa} with a maximal 
$\mu-\tau$ mixing angle, obtained through a non-orthogonal 
transformation, does not correspond to the present model.}, 
the large atmospheric angle must come from the neutrino sector.

An effective light neutrino mass matrix would transform as the conjugate of 
$\mathbf{S}$ under $SU(3) \otimes SU(3)$. If  the corresponding field is included, 
with the cutoff at $M_{\mathrm{Pl}}$, the  resulting (degenerate) neutrino masses 
are at most $O(v^{2}/M_{\mathrm{Pl}})$,  much smaller than the measured 
mass differences. One needs a model for neutrino masses and the natural 
choice in this GUT context is the seesaw mechanism with three $SU(5)$ 
singlets and their Majorana mass matrix. The obvious extension of the 
flavour symmetry is  $SU(3) \otimes  SU(3) \otimes SU(3)$ 
with $\mathbf{P}$, $\mathbf{F}$, $\mathbf{G}$ in 
$(\underline{1},\,  \underline{\bar{3}} ,\, \underline{\bar{3}})$ and 
additional breaking through Higgs fields in the bi-fundamental representation 
$(\underline{\bar{3}},\,  \underline{\bar{3}} ,\, \underline{1})$. 
The complete building of such a model is beyond the scopes of this letter, 
but it is worth noticing a nice feature of the seesaw 
mechanism in the present context. It comprises an $SU(5)$ invariant 
Majorana mass $\mathbf{M}_{\mathrm{R}}$ in the representation 
$(\underline{\bar{3}}\otimes \underline{\bar{3}},\, \underline{1},
\, \underline{1})$ of the flavour group and a Dirac mass  
$\mathbf{m}_{\mathrm{D}}$ in the 
$(\underline{\bar{3}},\, \underline{\bar{3}},\, \underline{1})$. The effective 
neutrino mass matrix is given (in the flavour basis) by the seesaw expression:
\begin{equation}
\mathbf{m}_{\nu} = \mathbf{m}_{\mathrm{D}}^T\mathbf{M}^{-1}_{\mathrm{R}}
\mathbf{m}_{\mathrm{D}} ~~.\label{mnu}
\end{equation}
By analogy with the quark case, we would expect a hierarchical structure in 
$\mathbf{m}_{\mathrm{D}}$ while $\mathbf{M}_{\mathrm{R}}=
\mathbf{m}_{\mathrm{D}} \circ \mathbf{m}_{\mathrm{D}}^T$ follows from the 
flavour symmetry, implying a quadratically stronger hierarchy.
The resulting hierarchy in $\mathbf{m}_{\nu}$ in general is much milder, 
although the precise relations depend on the structure of the field matrices.
The present model provides a non-abelian explanation 
for hierarchy compensation in the seesaw mechanism. 

For the sake of example, we assume for $\mathbf{m}_{\mathrm{D}}$ and  
$\mathbf{M}_{\mathrm{R}}$ 
textures analogous to those introduced in the quark sector, 
$\mathbf{m}_{\mathrm{down}}^T$ and $\mathbf{m}_{\mathrm{up}}$
in (\ref{textures}), respectively, and choose the 
$O(1)$ parameters to fit the experimental data. 
Although the textures generically predict a large atmospheric mixing 
as well as a very mild hierarchy among the neutrino mass eigenvalues,
some tuning of these parameters is needed to reproduce the data.
We obtain the following set:
$a=1.62,\,  b=1.01,\, c=.27,\,  g=1.43,\,  f=.17$. 

\vskip 12pt
\underline{\textit{Final Remarks}}

The present model, based on bottom-up flavour model building, is successful 
in describing  fermion masses and mixings, in explaining the hierarchies of up 
and down quarks, in exploiting the GUT breaking. It is natural in the technical 
sense that the mass matrices are defined by the breaking of flavour symmetries
through a set of fields.  Their configuration - direct product of $SU(3)$ group factors, 
matter in fundamental representations,  flavour symmetry breaking fields in 
bi-fundamental ones - reminds several setups in various frameworks. The 
breaking of one $SU(3)$ into its $SO(3)$ subgoup requires an additional
field in the symmetric representation and is a main ingredient in the realization
of the basic relation, $\mathbf{m}_{\mathrm{up}} = 
\mathbf{m}^{T}_{\mathrm{down}}\circ \mathbf{m}_{\mathrm{down}}$. 
Supersymmetric $SU(5)$ grand-unification plays a crucial role in the set-up of the 
model, and provides one of the two small parameters defining the flavour symmetry 
breaking scale. The fact that the t-quark coupling to one Higgs field is $O(1)$ 
implies that the first  flavour symmetry breaking occurs close to the cutoff scale 
$M_{\mathrm{Pl}}$ and that $\tan\beta$ has to be large.  

Work is still in progress on the construction of a detailed extension of the 
model to the seesaw mechanism, when both the Dirac and Majorana mass 
matrices are controlled by the breaking of the $SU(3)$ flavour symmetry 
associated to the heavy neutrinos. Crucial issues such as proton decay 
must also be addressed. Since the flavour and gauge symmetry breakings are
fixed (and related) one can tackle with some predictivity the supersymmetric
flavour problem so providing further tests of the model. Hopefully the features of the 
present model that might appear as weaknesses - in particular, CP violation - could 
be improved  through the choice of other assumptions and other frameworks, 
still guided by the relation $\mathbf{m}_{\mathrm{up}} = 
\mathbf{m}^{T}_{\mathrm{down}}\circ \mathbf{m}_{\mathrm{down}}$.

\vskip 12pt
\underline{\textit{Acknowledgements}} - This project is partially supported by the 
RTN European Program MRTN-CT-2004-503369. The work of I.M. was partially
supported by the University of Ferrara and C.A.S. is partially supported by the
ANR program ANR-05-BLAN-0193-02. The authors thank the Galileo Galilei
Institute for Theoretical Physics for the kind hospitality during the completion of
this work. We thank Z. Berezhiani for useful comments.

\end{document}